\documentclass[runningheads,a4paper]{llncs}

\usepackage{mathptmx}       
\usepackage{helvet}         
\usepackage{courier}        
\usepackage{type1cm}        
\usepackage{makeidx}         
\usepackage{graphicx}        
\usepackage[bottom]{footmisc}
\makeindex             
\usepackage{lscape} 
\usepackage{subfigure} 
\newcommand\apjl{ApJL }     
\newcommand\apjs{ApJS }
\newcommand\aap{A\&A }
\DeclareRobustCommand\ion[2]{%
  \mbox{#1\kern0.2em%
  \smaller\rmfamily%
  \edef\@tempa{\@car#2\@nil}%
  \ifcat1\@tempa%
  \@Roman{#2}%
  \else%
  \uppercase{#2}%
  \fi}}
  \newcounter{IonStage}

\def\ion#1#2{\ensuremath{\mathrm{#1}\;}{\protect\small\rm{#2}}}

\begin{document}

\title{Algorithm for removing secondary lines blended with Balmer lines in synthetic spectra of massive stars.}
\titlerunning{Removing secondary lines blended with Balmer lines.}
\author{Celia Rosa Fierro-Santill\'an\inst{1}, Jaime Klapp\inst{1}, Leonardo Di G. Sigalotti\inst{2}, Janos Zsarg\'o\inst{3}} 
  \institute{Instituto Nacional de Investigaciones Nucleares (ININ), Estado de M\'exico, M\'{e}xico,\\
        \and
            \'Area de F\'isica de Procesos Irreversibles, Departamento de Ciencias B\'asicas,\\
            Universidad Aut\'onoma Metropolitana-Azcapotzalco (UAM-A), Ciudad de M\'{e}xico, M\'{e}xico.\\
        \and
            Departamento de F\'{i}sica, Escuela Superior de F\'{i}sica y Matem\'aticas,\\
            Instituto Polit\'ecnico Nacional, Ciudad de M\'{e}xico, M\'{e}xico,\\ 
 \email{celia.fierro.estrellas@gmail.com},
 \email{jaime.klapp@inin.gob.mx},
 \email{leonardo.sigalotti@gmail.com}  
 \email{jzsargo@esfm.ipn.mx},
}

\authorrunning{Fierro-Santill\'an et al.}
\thispagestyle{empty} \maketitle \thispagestyle{empty}
\setcounter{page}{163}

\abstract{In order to measure automatically the equivalent width of the Balmer lines in 
a database of 40,000 atmosphere models, we have developed a program that mimics the work 
of an astronomer  in terms of identifying and eliminating secondary spectral lines mixed 
with the Balmer lines. The equivalent widths measured have average errors of 5 percent, 
which makes them very reliable. As part of the FIT\textit{spec} code, this program improves the 
automatic adjustment of an atmosphere model to the observed spectrum of a massive star.}


\section{Introduction}
\label{sec:intro}
The main restriction when studying astronomical objects is the impossibility of directly 
experiencing them. The massive stars have a period of evolution characteristic of millions 
of years and temperatures of the order of $10^4$\,K. Complex phenomena occurring in the atmosphere 
of the star can be simulated by a numerical code. In recent decades, there have been developed 
sophisticated stellar atmosphere codes such as TLUSTY \cite{Hubeny1995}, 
FASTWIND \cite{Santolaya1997}, \cite{Puls2005}, CMFGEN \cite{Hillier1998}, and the Postdam Wolf–Rayet 
(PoWR) code \cite{Grafener2002}, \cite{Hamann2003}, \cite{Sander2015}. As a result, significant 
advances have been achieved toward understanding the physical conditions prevailing in the 
atmospheres and winds of massive stars. 

The number of models generated to study an object grows exponentially depending on the number 
of parameters included in the simulation, with the consequent microprocessor time consumption. 
A strategy to address this problem is to generate a grid of models, covering characteristic 
values for each parameter, which can be used as a tool to study not only one, but an infinity 
of objects. With the use of the ABACUS-I supercomputer of the ABACUS Centre for Applied 
Mathematics and High Performance Computing of CINVESTAV (Mexico), it has recently been 
generated a grid with such characteristics \cite{Zsargo2016}. This grid covers a six-dimensional space 
with different values of the main parameters of the star, wind, and chemical composition. 
Currently the grid has 40,000 models of stellar atmospheres, and hence it would be impossible 
to compare \textit{by eye} the observed spectrum of a star with all models in the database.

In particular, the FIT\textit{spec} code \cite{Fierro2018} is a tool for the automatic fitting of synthetic 
stellar spectra. To adjust the effective temperature, FIT\textit{spec} requires as input the equivalent 
width (EW) of five helium lines: He II $\lambda\lambda$4541, 4200; He I $\lambda\lambda$4471, 4387, 4144; 
and He I + He II $\lambda$4026. Additionally, to adjust the surface gravity, the program requires 
the EW of six Balmer lines: H$_\beta$ $\lambda$4861, H$_\gamma$ $\lambda$4341, H$_\delta$ $\lambda$4102, 
H$_\epsilon$ $\lambda$, H$_\zeta$ $\lambda$3889, and H$_\eta$ $\lambda$3835. In order to achieve a good fit, 
it is important that the measurement of the EWs be as accurate as possible. The EW measured 
automatically may differ from what a human being would measure manually. It is important to 
reduce the effect of the lines mixed with the main line, since it overestimates the EW. In 
this paper we present a numerical method that reduces the effect of the mixed lines on the EW values.

\section{Measurement of the equivalent width by elimination of secondary lines}
\label{sec:measurement_ew}
\begin{figure*}
\centering
\includegraphics[width=0.7\linewidth]{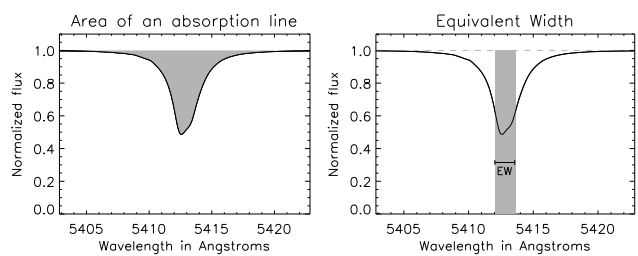}
\caption{Equivalent width of a spectral line.}
\label{f:ew}
\end{figure*}

The equivalent width (EW) is defined as the width of a rectangle with an area equal to the 
spectral line and a height equal to the continuum. For an experienced astronomer it is easy 
to identify by eye the initial ($w_i$) and final ($w_f$) wavelengths, as well as the continuous 
wavelength in order to measure the area of the spectral line and establish the EW (Fig.~\ref{f:ew}). 
However, a computer cannot identify these values directly, and to determine them we analyze 
a sample of 20 spectra of the database. The selection was made using random numbers, 
corresponding to the spectrum number. For each spectrum of the sample, the $w_i$ and $w_f$ 
values of the six Balmer lines considered were established by simply analyzing the 
spectrum by eye. The mean and the standard deviation were obtained and $w_i$ and $w_f$ were 
assumed as the values of the mean plus the standard deviation (Table~\ref{tab:lambda}). T
he value of the continuum was fixed at 1.0 since it corresponds to the  normalized spectra.
%
%
%

\begin{table}
\begin{center}
\caption{The $w_i$ and $w_f$ assumed are given by the sum of the mean plus the standard deviation 
from the 20 spectra of the sample.
\label{tab:lambda}}
\begin{tabular}{llllllll}
\multicolumn{2}{c}{} &\multicolumn{3}{c}{$w_i$ ($\AA$)} &
\multicolumn{3}{c}{$w_f$ ($\AA$)}\\
\multicolumn{1}{c}{Line} &\multicolumn{1}{c}{$\lambda_0$} &
\multicolumn{1}{c}{$\mu$} &\multicolumn{1}{c}{$\sigma$}&\multicolumn{1}{l}{Assumed}& 
\multicolumn{1}{c}{$\mu$} &\multicolumn{1}{c}{$\sigma$}&\multicolumn{1}{l}{Assumed}\\
\hline
H$_\beta$   & 4861.28~~~~ & ~4847.55~ & 4.61 & ~4842.94~~~~ & ~4877.60~ & 4.51 & ~4882.11\\
H$_\gamma$  & 4349.47~~~~ & ~4326.53~ & 3.47 & ~4323.06~~~~ & ~4356.51~ & 4.05 & ~4360.56\\
H$_\delta$  & 4101.71~~~~ & ~4086.11~ & 5.30 & ~4080.81~~~~ & ~4119.21~ & 6.00 & ~4125.22\\
H$_\epsilon$& 3970.08~~~~ & ~3959.26~ & 2.82 & ~3956.44~~~~ & ~3983.07~ & 4.31 & ~3987.37\\
H$_\zeta$   & 3889.02~~~~ & ~3878.75~ & 3.00 & ~3875.75~~~~ & ~3902.25~ & 3.49 & ~3905.74\\
H$_\eta$    & 3835.40~~~~ & ~3825.29~ & 2.71 & ~3822.58~~~~ & ~3847.60~ & 3.08 & ~3850.68\\
\hline
\end{tabular}
\end{center}
\end{table}

To determine if there are more than one spectral line between $w_i$ and $w_f$, 
we use Bolzano's theorem:
\textit{Let f be a continuous real function in a closed interval $[a, b]$ with 
$f(a)$ and $f(b)$ of opposite signs. Then there is at least one point c 
of the open interval $(a, b)$ with $f(c)=0$.}

This implies that when $f(a)$ and $f(b)$ have opposite signs, the function crosses 
the horizontal axis. We take advantage of this property to determine how many 
secondary lines are mixed with the main line that we want to measure. If the horizontal 
axis is arbitrarily moved and placed at an intermediate level between the continuum 
and the depth of the spectral line (Fig.~\ref{f:ew}), then the modified flow is obtained
\begin{equation}
\label{e:fm}
f_m= f -rl,
\end{equation}
where $f_m$ is the modified flow, $f$ is the normalized flow, and $rl$ is the reference
level, which can take any value between the continuum and the depth of the line.
We further assume that $f_m$ is a continuous function of the wavelength (w) in the
interval $[w_i, w_f]$, while $a$ and $b$ are two subsequent values of the wavelength. If
$f_m(a)$ and $f_m(b)$ have opposite signs, then there is a point c where $f_m(c)=0$. This
means that $f_m$ crosses the reference level. In Fig. 2, it is seen that the number of
spectral lines within the interval is given by the number of times(nc) that $f_m$ crosses
the reference line divided by two. In this way, the number of secondary lines $nsl$
mixed with a Balmer line is given by
\begin{equation}
\label{e:nsl}
nsl= \frac{nc-2}{2}.
\end{equation}
%
%
\begin{figure*}
\centering
\includegraphics[width=0.35\linewidth]{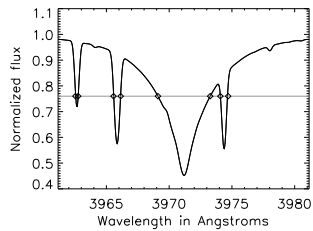}
\caption[width=0.8\linewidth]{The number of times that $f_m$ crosses the reference line (rl) in the interval $[w_i, w_f]$ 
divided by two corresponds to the number of spectral lines crossing the reference level.}
\label{f:reference_level}
\end{figure*}

\section{Algorithm}
\label{sec:algorithm}

The simplified version for a spectral line is shown. Twenty-five reference levels
were established iteratively.

\rule{10.0 cm}{0.02cm}\\
\textbf{input:} $f$ = normalized flux, float type array.\\
\hspace*{0.8 cm} w = wavelength, float type array.\\
\hspace*{0.8 cm} $lc$ = central wavelength of the spectral line, float.\\
\textbf{Output:} $f_m$ = normalized flux with secondary lines removed\\
\rule{10.0 cm}{0.02cm}\\
\textbf{Begin}\\
\hspace*{0.8 cm} fc = flux in the central wavelength of the spectral line\\
\hspace*{0.8 cm} step = (1 - fc)/25.0)\\
\hspace*{0.8 cm} level = 1\\
\hspace*{0.8 cm} \textbf{while} level less or equal to 25\\
\hspace*{1.5 cm} $rl$ = 1.0 – step * level\\
\hspace*{1.5 cm} n = number of elements of $f_m$\\
\hspace*{1.5 cm} $f_m$ = f - rl\\
\hspace*{1.5 cm} \textbf{for} i = 1 to n - 1\\
\hspace*{2.0 cm} sign = $f_m$[i] * $f_m$[i+1]\\
\hspace*{1.5 cm} \textbf{end for}\\
\hspace*{1.5 cm} nl = (number of elements of sign < 0)/2\\
\hspace*{1.5 cm} \textbf{if} nl less or equal to 1 then there are not secondary lines\\
\hspace*{1.5 cm} \textbf{if} nl > 1 then there are nl - 1 secondary lines\\
\hspace*{2.0 cm} \textbf{for} i=1 to nl -1\\
\hspace*{2.5 cm} $f_m$ in secondary line[i] = average flux of the rl\\
\hspace*{2.0 cm} \textbf{end for}\\
\hspace*{1.5 cm} \textbf{end if}\\
\hspace*{1.5 cm} level = level + 1\\
\hspace*{0.8 cm} \textbf{end while}\\
\hspace*{0.8 cm} $f_m$ = $f_m$ + rl\\
\hspace*{0.8 cm} return $f_m$\\
\textbf{end}\\
\rule{10.0 cm}{0.02cm}\\

\section{Errors}
\label{sec:errors}
The purpose of the algorithm is to measure the equivalent width of the Balmer
lines automatically, replacing the work of an experienced astronomer. Ideally, the
algorithm should obtain the same EW values as the astronomer. Assuming that the
error is the difference between both values and that the true value is the one that is
measured manually, we can calculate the error as:
\begin{equation}
\label{e:error}
error= \frac{EW_{auto}-EW_{man}}{EW_{man}}.
\end{equation}
\section{Results and discussion}
\label{sec:results}
\begin{figure*}
\centering
\includegraphics[width=0.9\textwidth]{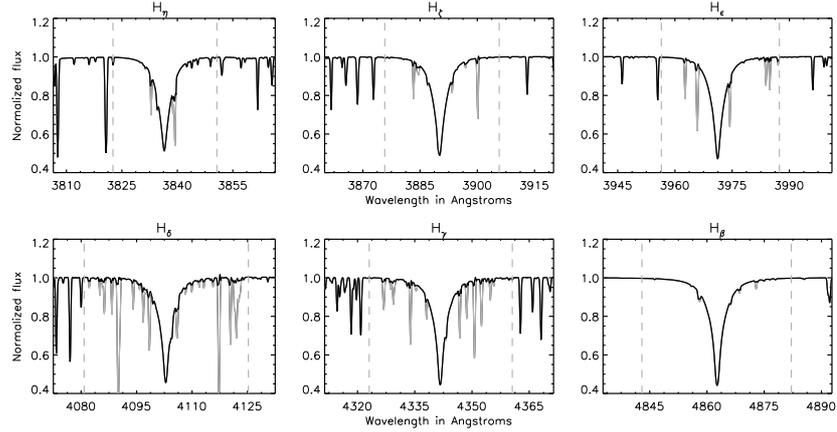}
\caption{Secondary lines removed (continuous gray line) before measuring the EW of the 
Balmer lines. Dashed gray lines indicate the $w_i$ and $w_f$ assumed when 
measuring the EW in all the spectra of the database.}
\label{f:eliminated}
\end{figure*}

Figure \ref{f:eliminated} shows a spectrum in which the secondary lines were removed with the use
of the algorithm. It is clearly seen that the program has identified and eliminated
the secondary lines properly. An astronomer can identify at first sight whether a
spectral line is isolated or there are several lines mixed. However, this is not a trivial
task for a computer. The term artificial intelligence is applied when a machine imitates 
some cognitive functions of human beings \cite{Rusell2009}. In this case, the algorithm
mimics the process of perceiving the spectral lines through the sense of sight.

\begin{figure*}
\centering
\includegraphics[width=0.7\textwidth]{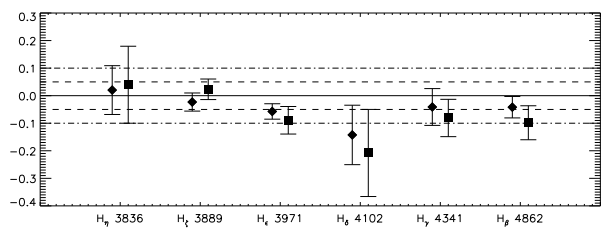}
\caption{Comparison of errors. The squares correspond to the errors in the EWs with secondary lines, while the
diamonds are the errors when the secondary lines are eliminated before measuring the EWs.}
\label{f:comp_errors}
\end{figure*}

Using equation~(\ref{e:error}), we calculate the errors for the six Balmer lines in 
the twenty spectra of the sample. Subsequently, the standard deviation and the average 
of the errors in each line are calculated, considering two cases when the secondary lines
are preserved or removed. The results are summarized in Table~\ref{tab:EW_balmer}, 
Additionally, Fig.~\ref{f:comp_errors} shows a comparison between the errors produced 
for each line in both cases.

It would be expected that when the EW is calculated by preserving the secondary
lines, its values would be overestimated. On the other hand, when the secondary
lines are removed, the overestimation will decrease, and even so the EWs would be
underestimated. However, Table~\ref{tab:EW_balmer} shows that in both cases, 
the EWs are underestimated in all the lines, except in H$_\eta$. This behavior is due 
to the $w_i$ and $w_f$ values having been considered in each case.

\begin{table}
\begin{center}
\caption{EW of measured Balmer lines 
\label{tab:EW_balmer}}
\begin{tabular}{llllll}
\multicolumn{2}{c}{} &\multicolumn{2}{c}{With secondary lines} &
\multicolumn{2}{c}{Without secondary lines}\\
\multicolumn{1}{c}{line} & \multicolumn{1}{c}{$\lambda_0$} &
\multicolumn{1}{c}{~~~~~Average error} & \multicolumn{1}{c}{$\sigma$} &
\multicolumn{1}{c}{~~~~~Average error} & \multicolumn{1}{c}{$\sigma$} \\
\hline
H$\beta$   & 4861.28 & ~~~~~~~~~~-0.0920 & 0.0838 & ~~~~~~~~~~-0.0402 & 0.0415 \\
H$\gamma$  & 4349.47 & ~~~~~~~~~~-0.1147 & 0.1694 & ~~~~~~~~~~-0.0953 & 0.2484 \\
H$\delta$  & 4101.71 & ~~~~~~~~~~-0.1833 & 0.1881 & ~~~~~~~~~~-0.1560 & 0.1549 \\
H$\epsilon$& 3970.08 & ~~~~~~~~~~-0.0837 & 0.0560 & ~~~~~~~~~~-0.0577 & 0.0299 \\
H$\zeta$   & 3889.02 & ~~~~~~~~~~~0.0525 & 0.1230 & ~~~~~~~~~~-0.0022 & 0.0886 \\
H$\eta$    & 3835.40 & ~~~~~~~~~~~0.1183 & 0.3405 & ~~~~~~~~~~~0.0427 & 0.1364 \\
\hline
\end{tabular}
\end{center}
\end{table}

When the EWs are measured by keeping the secondary lines, the values of $w_i$ 
and $w_f$ are established closer to the central wavelength to avoid the effect of such lines.
This method underestimates the EWs, especially in those spectra where the Balmer
lines are broadened by gravitational effects. On the other hand, the algorithm that
suppresses the secondary lines, allowed to fix the values of $w_i$ and $w_f$ 
more realistically, from the spectra of the sample. This algorithm obtains values closer to those
that an astronomer would measure regardless of whether the Balmer lines are narrow or broadened.

Figure~\ref{f:comp_errors} shows that when measuring the EWs while retaining the secondary lines,
the average errors are 10 percent, while when eliminating the secondary lines, the
average errors are reduced to 5 percent. Only for H$_\delta$ the errors are greater than 15
percent. As it can be seen in Figure~\ref{f:eliminated}, this line presents a large number of mixed lines.

\section{Conclusions}
\label{sec:conclusions}

Properly measuring the EWs of a spectrum is a task that requires an experienced
astronomer. However, the time required grows proportionally to the number of
spectra and the lines measured in each spectrum. Obtaining the value of the EWs
in a database of 40,000 spectra in a reasonable time is an impossible task to perform
manually. In this work, we have presented an algorithm, which has the ability to
perform a realistic measurement, identify the secondary lines mixed with the Balmer lines, 
and then eliminate them before calculating the EW, in a way similar to what an 
experienced astronomer would do.

The algorithm improves the results obtained in the previous version of FIT\textit{spec},
reducing the error from 10 to 5 percent. By improving the value of EWs, the algorithm 
also allows to increase the quality of the automatic adjustment of spectra
made by FIT\textit{spec}.

\section*{Acknowledgments}
The authors acknowledge the use of the ABACUS-I supercomputer at the Laboratory of 
Applied Mathematics and High-Performance Computing of the Mathematics Department of 
CINVESTAV-IPN, where this work was performed. The research leading to these results 
has received funding from the European Union's Horizon 2020 Programme under the 
ENERXICO Project, grant agreement no 828947 and under the Mexican 
CONACYT-SENER-Hidrocarburos grant agreement B-S-69926. J. K. acknowledges financial 
support by the Consejo Nacional de Ciencia y Tecnolog\'ia (CONACyT), M\'exico, 
under grant 283151.

\bibliography{ms}

\begin{thebibliography}{10}
\providecommand{\url}[1]{\texttt{#1}}
\providecommand{\urlprefix}{URL }

\bibitem{Hubeny1995}
{Hubeny}, I. \& {Lanz}, T.: {Non-LTE line-blanketed model atmospheres of hot stars.
1: Hybrid complete linearization/accelerated lambda iteration method}. \apj 439, 875-904 (1995).

\bibitem{Santolaya1997}
{Santolaya-Rey}, A. E., {Puls}, J., \& {Herrero}, A.: {Atmospheric NLTE-models for the
spectroscopic analysis of luminous blue stars with winds}. \aap 323, 488-512 (1997).

\bibitem{Puls2005}
{Puls}, J., {Urbaneja}, M. A., {Venero}, R., {Repolust}, T., {Springmann}, U., {Jokuthy}, A., \& {Mokiem}, M. R.: {Atmospheric NLTE-models for the spectroscopic analysis of blue stars with winds-II. Line-blanketed models}. \aap 435(2), 669-698 (2005).

\bibitem{Hillier1998}
{Hillier}, D. J., \& {Miller}, D. L.: {The treatment of non-LTE line blanketing in spherically expanding outflows}. \apj 496(1), 407 (1998).

\bibitem{Grafener2002}
{Gr\"afener}, G., {Koesterke}, L., \& {Hamann}, W. R.: {Line-blanketed model atmospheres
for WR stars}. \aap 387(1), 244-257 (2002).

\bibitem{Hamann2003}
{Hamann}, W. R., \& {Gr\"afener}, G.: {A temperature correction method for expanding
atmospheres}. \aap 410(3), 993-1000 (2003).

\bibitem{Sander2015}
{Sander}, A., {Shenar}, T., {Hainich}, R., {Gímenez-García}, A., {Todt}, H., \& {Hamann}, W. R.: {On the consistent treatment of the quasi-hydrostatic layers in hot star atmospheres}. \aap 577, A13 (2015).

\bibitem{Zsargo2016}
{Zsarg\'o}, J., {Fierro}, C. R., {Klapp}, J., {Arrieta}, A., {Arias}, L., \& {Hillier}, D. J. Database of CMFGEN Models in a 6-Dimensional Space. In Latin American High Perfor-
mance Computing Conference (pp. 387-392). Springer, Cham.(2016, August).

\bibitem{Fierro2018}
{Fierro-Santillán}, C. R., {Zsarg\'o}, J., {Klapp}, J., {D\'iaz-Azuara}, S. A., {Arrieta}, A., {Arias}, L., \& {Sigalotti}, L. D. G.: {FITspec: A New Algorithm for the Automated Fit of Synthetic Stellar Spectra for OB Stars}. \apjs 236(2), 38 (2018).

\bibitem{Rusell2009}
{Russell}, S. J., \& {Norvig}, P.: {Artificial intelligence: a modern approach. Malaysia}. Pearson Education Limited (2016).

\end{thebibliography}
\bibliographystyle{splncs03}

\end{document}